\documentclass[aps,prd,10pt,notitlepage,nofootinbib,superscriptaddress,showkeys,showpacs]{revtex4-2}
\pdfoutput=1
\usepackage{amstext,amsmath,amssymb,amsfonts,bbm,euscript,color,dsfont}
\usepackage{hyperref}
\usepackage{amsthm}
\usepackage{color}
\usepackage{multirow}

\usepackage{verbatim}

\usepackage{graphicx}

\usepackage{latexsym}

\newcommand{\bea}{\begin{eqnarray}}	
\newcommand{\eea}{\end{eqnarray}}
\newcommand{\be}{\begin{equation}}	
\newcommand{\ee}{\end{equation}}
\newcommand{\beq}{\begin{equation}}	
\newcommand{\eeq}{\end{equation}}

\newcommand{\Z}{{\mathbb Z}}
\newcommand{\C}{{\mathbb C}}

\def\R{\relax\ifmmode {\mathbb R}  \else${\mathbb R}$\fi}
\def\C{\relax\ifmmode {\mathbb C}  \else${\mathbb C}$\fi}
\def\Z{\relax\ifmmode {\mathbb Z}  \else${\mathbb Z}$\fi}
\def\N{\relax\ifmmode {\mathbb N}  \else${\mathbb N}$\fi}
\def\I{\relax\ifmmode {\mathbb I}  \else${\mathbb I}$\fi}

\begin{document}

\title{Interpolating gauge-fixing for Yang-Mills-Chern-Simons theory in $D=3$}  


\author{Daniel O. R. Azevedo}\email{daniel.azevedo@ufv.br}
\affiliation{Departamento de F\'isica, Universidade Federal de Vi\c cosa,
	Campus Universit\'ario, Avenida Peter Henry Rolfs s/n, 36570-900, Vi\c cosa, MG, Brazil}
 \affiliation{Ibitipoca Institute of Physics (IbitiPhys), Concei\c c\~ao do Ibitipoca, 36140-000, MG, Brazil}
\author{Oswaldo M. Del Cima} \email{oswaldo.delcima@ufv.br}
\affiliation{Departamento de F\'isica, Universidade Federal de Vi\c cosa,
	Campus Universit\'ario, Avenida Peter Henry Rolfs s/n, 36570-900, Vi\c cosa, MG, Brazil}
 \affiliation{Ibitipoca Institute of Physics (IbitiPhys), Concei\c c\~ao do Ibitipoca, 36140-000, MG, Brazil}
\author{Thadeu S. Dias} \email{thadeu.dias@ufv.br}
\affiliation{Departamento de F\'isica, Universidade Federal de Vi\c cosa,
	Campus Universit\'ario, Avenida Peter Henry Rolfs s/n, 36570-900, Vi\c cosa, MG, Brazil}
 \affiliation{Ibitipoca Institute of Physics (IbitiPhys), Concei\c c\~ao do Ibitipoca, 36140-000, MG, Brazil}
\author{Em\'ilio D. Pereira} \email{emilio.drumond@ufv.br}
\affiliation{Departamento de F\'isica, Universidade Federal de Vi\c cosa,
	Campus Universit\'ario, Avenida Peter Henry Rolfs s/n, 36570-900, Vi\c cosa, MG, Brazil}
 \affiliation{Ibitipoca Institute of Physics (IbitiPhys), Concei\c c\~ao do Ibitipoca, 36140-000, MG, Brazil}


\begin{abstract}
The Yang-Mills-Chern-Simons theory in three-dimensional Minkowski space-time is studied in a gauge-fixing scheme which interpolates between the covariant gauge and light-cone gauge, the interpolating gauge-fixing. The ultraviolet finiteness of the theory is proved via the Becchi-Rouet-Stora (BRS) algebraic renormalization procedure, which allows us to demonstrate the vanishing of all $\beta$-functions and all anomalous dimensions to all orders in perturbation theory.

 

\end{abstract}

\maketitle

{\it Dedicated to the memory of Prof. Manfred Schweda (1939-2017).}
\\

\section{Introduction}
It is a well established fact that Yang-Mills-Chern-Simons (YMCS) in $D=3$ dimensions is a finite theory \cite{giavarini,cfhp3}, meaning that no parameter suffer from radiative corrections coming from loop computations. This result was first obtained by explicit computations up to two loops \cite{pisarski,deser2,martin,giavarini}, as well as the $N=1$ super-YMCS theory \cite{ruiz} and a partial proof for the $N=2$ super-YMCS theory in the Wess-Zumino gauge \cite{maggiore}. Later, it was proved algebraically to all orders in perturbation theory \cite{cfhp3}, in the Landau gauge, which presents series of special features compared to a general linear covariant gauge, but can easily be extended to those. The main advantage of the algebraic method \cite{becchi,piguet1,piguet2} is the independence of a regularization scheme, which could be difficult to implement in an invariant manner due to the presence of a Levi-Civita tensor. 

Another possibility is the use of non-covariant gauges, such as the light-cone gauge, which presents some advantages in being adopted. Among them, we have the decoupling of the ghost field from the gauge field, the presence of only transverse modes and its high application in supersymmetric models \cite{noncovariant.book}. On the other hand, other difficulties rise up, for example the quantization of gauge theories \cite{noncovariant}, due to nonlocal divergences coming from the nonpolynomial dependence on the external momenta, which does not allow to naively use general results of the power-counting theorem. One way to circumvent those difficulties is to use a gauge-fixing which interpolates between the Landau gauge and the light-cone gauge with a Leibbrandt-Mandelstam prescription \cite{piguet3,landsteiner}. This is done by introducing a gauge parameter $\zeta$ defined on the positive real line, whose limiting values recover the linear covariant gauge ($\zeta=0$) and light-cone gauge ($\zeta=\infty$). Moreover, the use of such gauge-fixing allows us to use general results of renormalization theory \cite{lowenstein,zimmermann}, since the propagators are compatible with the power-counting theorem for finite values of $\zeta$, leading to only local couneterterms being needed to renormalize the theory.

In this letter, we make use of such interpolating gauge to study the algebraic renormalization of YMCS theory at all orders in radiative corrections. The work is organized as follows: in Section II, we introduce the YMCS action with the interpolating gauge-fixing and analyze the gluon propagator; in Section III we introduce the tree-level symmetries obeyed by the action; in Section IV we prove the ultraviolet finiteness of the theory at all orders in perturbation.

\section{The YMCS action in the interpolating gauge}
The tree-level action is given by the sum of the invariant action $\Sigma_{inv}$ and the gauge-fixing action $\Sigma_{gf}$. The invariant action for the Yang-Mills-Chern-Simons theory is
\begin{equation}\label{sinv}
	\Sigma_{inv} = \int d^3x\, \text{Tr} \left[-\frac{1}{4}F_{\mu\nu}F^{\mu\nu}+\frac{1}{2}m\epsilon^{\mu\nu\rho}\left(A_\mu\partial_\nu A_\rho +\frac{2g}{3}A_\mu A_\nu A_\rho\right)\right],
\end{equation}
and the interpolating gauge-fixing action is
\begin{equation}\label{sgf}
	\begin{split}
		\Sigma_{gf} &= s \int d^3x\, \text{Tr}\left[K_-(\zeta ( n^{\ast\mu}\partial_\mu)D_0-\partial_\mu A^\mu) + \Bar{c}(D_0+n^\mu A_\mu)+\frac{1}{2}\alpha K_-K_0\right]\\
		&=  \int d^3x\, \text{Tr}\left[K_0(\zeta (n^{\ast\mu}\partial_\mu)D_0-\partial_\mu A^\mu) - K_-(\zeta (n^{\ast\mu}\partial_\mu)D_+ + \partial^\mu D_\mu c)  + b(D_0+n^\mu A_\mu) - \Bar{c}(D_+-n^\mu D_\mu c) + \frac{1}{2}\alpha K_0^2\right],
	\end{split}    
\end{equation}
where the BRS transformations are
\begin{equation}
	\begin{aligned}
		&sA_\mu = -D_\mu c, \quad sc=gc^2, \quad s\Bar{c}=b, \quad sb=0\\
		&sK_-=K_0, \quad sK_0=0, \quad sD_0=D_+, \quad sD_+=0,
	\end{aligned}
\end{equation}
where the subindices $(-,0,+)$ indicate the auxiliary fields ghost number and the covariant derivative of the ghost field $D_\mu c$ is
\begin{equation}
	D_\mu c = \partial_\mu c + g[A_\mu,c].
\end{equation}
All fields are Lie algebra valued $\phi=\phi^a T^a$, with $T^a$ being the generators of the gauge group, obeying $\text{Tr}(T^a T^b) = \delta^{ab}$. We assume a Minkowskian metric $\eta_{\mu\nu}=\text{diag}(+,-,-)$ and $n^\mu = (|\mathbf{n}|,\mathbf{n})$ as a light-like vector, with $n^{*\mu}=  (|\mathbf{n}|,-\mathbf{n})$ being the dual vector. In the above expressions, $g$ is the coupling constant of the $SU(N)$ gauge group and $m$ is the Chern-Simons topological mass parameter. It is important to note that the gauge-fixing action (\ref{sgf}) has two distinct gauge parameters, $\alpha$ and $\zeta$. The first one is the usual gauge parameter for the linear covariant gauge. The second, interpolates between the linear covariant gauge ($\zeta = 0$) and the light cone gauge ($\zeta \rightarrow \infty$).
This fact can be seen by the gluon propagator
\begin{equation}
\begin{split}
	\langle T A_\mu^a A_\nu^b \rangle & =  \frac{i\delta^{ab}}{\partial^2+m^2}\left[ \frac{\alpha}{(\partial^2_\zeta)^2}\frac{\partial^2(\partial^2+m^2)+m^2\zeta(n^{\ast\mu}\partial_\mu)(n^{\mu}\partial_\mu)}{\partial^2}\partial_\mu\partial_\nu \right.\\ & \left. + \eta_{\mu\nu} - \frac{\partial^2\partial_\mu\partial_\nu}{\partial^2_\zeta} - \frac{m}{\partial^2}\epsilon_{\mu\rho\nu}\partial^\rho -\frac{\zeta(n^{\ast\mu}\partial_\mu)}{\partial^2_\zeta}(\partial_\mu n_\nu + \partial_\nu n_\mu) \right],
\end{split}
\end{equation}
where $\partial^2_\zeta = \partial^2 + \zeta(n^{*\mu}\partial_\mu)(n^\mu \partial_\mu)$. By taking the appropriate values of $\zeta$, we recover the covariant propagator 
\begin{equation}
	\langle T A_\mu^a A_\nu^b \rangle|_{\zeta=0} =  \frac{i\delta^{ab}}{\partial^2+m^2}\left[\eta_{\mu\nu} - \frac{\partial_\mu\partial_\nu}{\partial^2} - \frac{m}{\partial^2}\epsilon_{\mu\rho\nu}\partial^\rho\right] +i\delta^{ab} \frac{\alpha}{\partial^2}\frac{\partial_\mu\partial_\nu}{\partial^2} ,
\end{equation}
and the light-cone propagator
\begin{equation}
	\langle T A_\mu^a A_\nu^b \rangle|_{\zeta\rightarrow \infty} =  \frac{i\delta^{ab}}{\partial^2+m^2}\left[\eta_{\mu\nu} - \frac{m}{\partial^2}\epsilon_{\mu\rho\nu}\partial^\rho -\frac{\partial_\mu n_\nu + \partial_\nu n_\mu}{n^{\mu}\partial_\mu} \right].
\end{equation}
It should be stressed that since the interpolating parameter $\zeta$ is a gauge parameter, it cannot enter in any physical correlation functions of the model. In addition, for any finite value of $\zeta$, the relevant propagator is compatible with the requirements of the power-counting theorem \cite{lowenstein,zimmermann}, we can make use of the quantum action principle and the algebraic renormalization procedure \cite{piguet1,piguet2} to study the perturbative ultraviolet behavior of the theory to all orders.
\section{The symmetries of the action}
The gauge-fixed Yang-Mills-Chern-Simons action, (\ref{sinv}) and (\ref{sgf}), is invariant under BRS transformations. To pursue the algebraic renormalization procedure, it is necessary to introduce external source fields to control the non-linear transformations \cite{becchi} of the gauge field $A_\mu$ and the Faddeev-Popov ghost $c$:
\begin{equation}\label{sources}
	\Sigma_{ext} = \int d^3x~\text{Tr}\left[\rho^\mu s A_\mu + \sigma s c\right] .
\end{equation} 

Therefore, the complete tree-level action reads:
\begin{equation}\label{gamma0}
	\Gamma^{(0)} = \Sigma_{inv} + \Sigma_{gf} + \Sigma_{ext},
\end{equation}
and it obeys the Slavnov-Taylor identity
\begin{equation}\label{STid}
	\mathcal{S}(\Gamma^{(0)}) = \int d^3x~\text{Tr}\left[\frac{\delta \Gamma^{(0)}}{\delta \rho^\mu}\frac{\delta \Gamma^{(0)}}{\delta A_\mu} + \frac{\delta \Gamma^{(0)}}{\delta \sigma}\frac{\delta \Gamma^{(0)}}{\delta c} + b\frac{\delta \Gamma^{(0)}}{\delta \bar{c}} +K_0\frac{\delta \Gamma^{(0)}}{\delta K_-} + D_+\frac{\delta \Gamma^{(0)}}{\delta D_0}\right] = 0,
\end{equation}
where it will be convenient for later use to define the linearized Slavnov-Taylor operator
\begin{equation}\label{STlin}
	\mathcal{S}_{\Gamma^{(0)}} = \int d^3x~\text{Tr}\left[\frac{\delta \Gamma^{(0)}}{\delta \rho^\mu}\frac{\delta }{\delta A_\mu}+ \frac{\delta \Gamma^{(0)}}{\delta A_\mu}\frac{\delta }{\delta \rho^\mu} + \frac{\delta \Gamma^{(0)}}{\delta \sigma}\frac{\delta }{\delta c}+ \frac{\delta \Gamma^{(0)}}{\delta c}\frac{\delta }{\delta \sigma} + b\frac{\delta }{\delta \bar{c}} +K_0\frac{\delta }{\delta K_-} + D_+\frac{\delta }{\delta D_0}\right].
\end{equation}
We also have four gauge conditions from the auxiliary Lagrange multiplier fields
\begin{equation}\label{gaugecond}
	\begin{aligned}
		&\frac{\delta \Gamma^{(0)}}{\delta b} = D + n^\mu A_\mu, \\
		&\frac{\delta \Gamma^{(0)}}{\delta K_0} = \zeta (n^{\ast\mu}\partial_\mu)D_0 - \partial_\mu A^\mu + \alpha K_0,\\
		&\frac{\delta \Gamma^{(0)}}{\delta D_0} = -\zeta (n^{\ast\mu}\partial_\mu)K_0 + b,\\
		&\frac{\delta \Gamma^{(0)}}{\delta D_+} = -\zeta (n^{\ast\mu}\partial_\mu)K_- + \bar{c},
	\end{aligned}
\end{equation}
as well as two ghost equations
\begin{equation}\label{ageq}
	\begin{aligned}
		&\frac{\delta \Gamma^{(0)}}{\delta \bar{c}} + n^\mu\frac{\delta \Gamma^{(0)}}{\delta \rho^\mu} = -D_+,\\
		&\frac{\delta \Gamma^{(0)}}{\delta K_-} - \partial^\mu\frac{\delta \Gamma^{(0)}}{\delta \rho^\mu} = -\zeta (n^{\ast\mu}\partial_\mu)D_+,
	\end{aligned}
\end{equation}
and an integrated anti-ghost equation
\begin{equation}\label{geq}
	\int d^3x \left\{ \frac{\delta \Gamma^{(0)}}{\delta c} + g\left[K_-,\frac{\delta \Gamma^{(0)}}{\delta K_0}\right] + g\left[D_0,\frac{\delta \Gamma^{(0)}}{\delta D_+}\right] + g\left[\bar{c},\frac{\delta \Gamma^{(0)}}{\delta b}\right] \right\} = g\int d^3x \left\{[c,\sigma] - [\rho^\mu,A_\mu] + \alpha[K_-,K_0] \right\}.
\end{equation}

\section{Renormalization}
The study of perturbative renormalization in the framework of the  algebraic method consists in the possibility of extending the symmetries of the classical action ($\Gamma^{(0)}$) to the quantum vertex functional $\Gamma$, also called quantum action in view of its formal perturbative expansion
\begin{equation}
	\Gamma = \Gamma^{(0)} + O(\hslash).
\end{equation}
Before discussing the quantum extension of the symmetries, we must make a remark about power-counting renormalizability. Owing to the fact that the Yang-Mills-Chern-Simons theory is  superrenormalizable, we have the following expression for the superficial degree of divergence of any one-particle irreducible Feynman diagram $\gamma$ \cite{maggiore,cfhp3}:
\begin{equation}
	d(\gamma) = 3 - \sum_{\Phi=\phi,g}d_\Phi N_\Phi,
\end{equation}
where $N_\phi$ is the number of external legs in $\gamma$ corresponding to the field $\phi$, $d_\phi$ is its dimension, $N_g$ is the power of the coupling constant appearing in the integral representation of $\gamma$, and $d_g$ is the dimension of the coupling constant. It is crucial to notice that the coupling constant $g$ is dimensionful {\it i.e.} $d_g= 1/2$, a key feature of superrenormalizable theories. 

In order to use the general results of the quantum action principle, we treat $g$ here as an external field. The ultraviolet (UV) dimensions of all fields are gathered at Table \ref{tab1}, as well as their ghost numbers.
\begin{table}
	\begin{tabular}{| c | c | c | c | c | c | c | c | c | c | c | c |}
		\hline
		Fields & $A_\mu$ & $b$ & $c$ & $\bar{c}$ & $K_0$ & $K_-$ & $D_0$ & $D_+$ & $\rho_\mu$ & $\sigma$ & $g$\\
		\hline
		UV Dimension & 1/2 & 5/2 & -1/2 & 5/2 & 3/2 & 3/2 & 1/2 & 1/2 & 5/2 & 7/2 & 1/2 \\
		\hline
		$\Phi\Pi$-charge & 0 & 0 & 1 & -1 & 0 & -1 & 0 & 1 & 1 & -2 & 0 \\
		\hline
	\end{tabular}
	\caption{Dimension and ghost numbers of the fields and coupling constant}
	\label{tab1}
\end{table}
From the inclusion of the coupling constant in the degree of divergence, it can be seen that any counterterm appearing in the quantum action, since they are generated by loop graphs, will be at least of power two in $g$. Consequently, the dimension of such counterterms shall be bounded by 2, thus by the same reason all possible breakings of the Slavnov-Taylor identity are also bounded by UV dimension 2.

We now focus on the extension of the symmetries to the quantum level, {\it i.e.} the study of possible anomalies and non-invariant counterterms. The gauge conditions (\ref{gaugecond}) and ghost equations (\ref{ageq}) can all be shown to hold to all orders in perturbation theory \cite{piguet2}, since they are broken linearly in the quantum fields. Moreover, the anti-ghost equation (\ref{geq}), due to the term $\alpha[K_-,K_0]$ in its breaking, cannot be extended to the quantum level since it is non-linear in the quantum fields, $K_-$ and $K_0$. However, the breaking of the anti-ghost equation might be linearized if we adopt the Landau gauge $(\alpha = 0)$; in this case, the anti-ghost equation can be extended to the quantum level. The use of the Landau gauge provides two main advantages: i) the nonrenormalization of the anti-ghost equation controls the dependence of the theory on the ghost $c$ \cite{blasi1}; ii) it also leads to another invariance, the rigid gauge transformations:
\begin{equation}\label{riginv}
	\mathcal{W}^a \Gamma^{(0)} = \int d^3x \sum_\phi f^{abc} \phi^b \frac{\delta \Gamma^{(0)}}{\delta \phi^c} = 0,
\end{equation}
where $\phi$ stands for any field in the action (\ref{gamma0}). Due to these advantages, we will assume the Landau gauge from now on. Therefore, all functional identities, (\ref{gaugecond}) and (\ref{ageq}), of the model have their extension at the quantum level, including the anti-ghost equation (\ref{geq}) and rigid symmetry (\ref{riginv}).

The extension of the Slavnov-Taylor identity (\ref{STid}) to the quantum action entails the solution of the cohomology problem of the linearized Slavnov-Taylor operator $\mathcal{S}_{\Gamma^{(0)}}$, defined by (\ref{STlin}), in the sector of ghost number 1 of local integrated polynomials in the quantum fields. Defining the possible quantum breaking of (\ref{STid}) at order $n$ in $\hbar$ $(\hbar^n)$ as
\begin{equation}
	\mathcal{S}(\Gamma) = \hslash^n \Delta + O(\hslash^{n+1}),
\end{equation}
we can see that the nilpotency of the Slavnov-Taylor operator $\mathcal{S}_{\Gamma^{(0)}}$, which comes from the nilpotency of the BRS-transformations ($s^2=0$), together with 
\begin{equation}
	\mathcal{S}_{\Gamma} = \mathcal{S}_{\Gamma^{(0)}} + O(\hslash),
\end{equation}
enforces the Wess-Zumino consistency condition
\begin{equation}\label{WZ}
	\mathcal{S}_{\Gamma^{(0)}}\Delta = 0,
\end{equation}
characterizing a cohomology problem in the sector of ghost number one. The solutions of (\ref{WZ}) is of the form
\begin{equation}
	\Delta = \Delta^{(1)} + \mathcal{S}_{\Gamma^{(0)}}\Delta^{(0)},
\end{equation}
where $\Delta^{(1)}$ and $\Delta^{(0)}$ are local integrated functionals in the quantum fields with ghost number 1 and 0, respectively. Only the non-trivial cocycle $\Delta^{(1)}$ can represent a true anomaly of the Slavnov-Taylor identity, since $\Delta^{(0)}$ is the so called non-invariant counterterms that should be absorbed by the quantum action order by order. However, it is a known fact that the cohomology sector of ghost number 1 is empty in three dimensions \cite{brandt,barnich1,barnich2,bandelloni}. Allied to this fact, as explained above, all possible breakings $\Delta$ are bounded by UV dimension 2, and since the non-trivial cocycle $\Delta^{(1)}$ is empty, thus the only solution is the trivial coboundary $\mathcal{S}_{\Gamma^{(0)}}\Delta^{(0)}$, meaning that the theory is free from any gauge anomalies.

Moving on to the sector of ghost number 0, the solutions to the Wess-Zumino consistency conditions are invariant counterterms of the theory, representing the renormalization of physical parameters, such as masses and coupling constants, and nonphysical ones, which are redefinitions of field amplitudes, which are the nontrivial cocycles and coboundaries, respectively \cite{piguet2,MShed}. This corresponds to the multiplicative renormalization analysis of the model, namely the stability of the tree-level action subjected to perturbation, $\Gamma^{(0)} + \varepsilon\Gamma^{ct}$, where $\Gamma^{ct}$ is the invariant counterterm build up by all local field monomials compatible with
the homogeneous versions of equations (\ref{gaugecond}-\ref{geq}) and (\ref{riginv}):
\begin{equation}
	\frac{\delta \Gamma^{ct}}{\delta b} = \frac{\delta \Gamma^{ct}}{\delta K_0} = \frac{\delta \Gamma^{ct}}{\delta D_0} = \frac{\delta \Gamma^{ct}}{\delta D_+} = \int d^3x \frac{\delta \Gamma^{ct}}{\delta c} = 0,
\end{equation}
which means that $\Gamma^{ct}$ does not depend on the fields $b$, $K_0$, $D_0$ and $D_+$, and the ghost can only enter as a derivative $\partial_\mu c$. Beyond that, the ghost equations (\ref{ageq}) restrict the fields $\bar{c}$, $K_-$ and $\rho^\mu$ to enter only through the combination $\bar{\rho}^\mu = \rho^\mu - n^\mu \bar{c} + \partial^\mu K_-$. It turns out that the most general counterterm action that obeys all constraints is the Chern-Simons action,
\begin{equation}
	\Gamma^{ct} = z_m \int d^3 \text{Tr} \left[m\epsilon^{\mu\nu\rho}\left(A_\mu\partial_\nu A_\rho +\frac{2g}{3}A_\mu A_\nu A_\rho\right)\right] = z_m m \frac{\partial}{\partial m}\Gamma^{(0)},
 \label{eq23}
\end{equation}
where $z_m$ is an arbitrary parameter. 

It follows from Eq.(\ref{eq23}) that only the Chern-Simons topological mass term can get contributions from radiative corrections. However, since the Chern-Simons action (\ref{eq23}) is only invariant under BRS transformations up to a total derivative, then not locally invariant \cite{barnich}, it was proved through the local version of the Callan-Zymanzik equation that the $\beta$-function associated with the mass parameter $m$ vanishes \cite{cfhp1,cfhp2,cfhp3}. Therefore, neither the parameters (coupling constant and mass parameter) nor the quantum fields get any radiative corrections, consequently their respective $\beta$-functions and anomalous dimensions are identically zero, which renders the ultraviolet finiteness of the YMCS theory in the interpolating gauge at all orders in perturbation theory.

\section{Conclusions}

In this letter, we analyze the quantum consistency of the Yang-Mills-Chern-Simons theory by adopting the interpolating gauge in the scope of the Becchi-Rouet-Stora algebraic renormalization method, which is independent on any regularization scheme. From the gluon propagator, we show that the interpolating parameter $\zeta$ is consistent with the asymptotic cases: linear covariant gauge ($\zeta = 0$) and light cone gauge ($\zeta \rightarrow \infty$). Moreover, we show that all functional identities are extended to the quantum level when the Landau gauge $(\alpha=0)$ is assumed, as well as the theory is free from any gauge anomalies. Furthermore, the multiplicative renormalizability of the model is proved, and as a byproduct of its superrenormalizability and the fact that the Chern-Simons action is only BRS invariant up to a total derivative, the $\beta$-functions associated to the coupling constants ($g$) and the Chern-Simons mass term ($m$), and the anomalous dimensions of all the quantum fields vanish. Finally, the YMCS theory in the interpolating gauge is ultraviolet finiteness at all orders in perturbation theory.

\subsection*{Acknowledgements}

The authors thank CAPES-Brazil for invaluable financial help. 


\end{document}